\documentclass[reprint, amsmath, amssymb, aps, pra, showkeys]{revtex4-2}
\usepackage{graphicx}
\usepackage{bm}
\usepackage{hyperref}

\begin{document}

\preprint{APS/123-QED}

\title{Quantum Correlations in One Parameter Mixed Quantum States}
\thanks{A footnote to the article title}

\author{Kapil K. Sharma} \email{iitbkapil@gmail.com, kapil.sharma3@sharda.ac.in} 
\affiliation{Dept. of Physics, Sharda School of Basic Sciences and Research, Greater Noida, Uttar Pradesh 201310, India}
\affiliation{Nex Gen Cyber Security and Quantum Systems Facility, Sharda Univ., Greater Noida, Uttar Pradesh 201310, India}

\author{Rishikant Rajdeepak}
\email{rishikant.rajdeepak@dypiu.ac.in}
\affiliation{D Y Patil International University, Akurdi, Pune, Maharashtra 411044, India}

\author{Fatih Ozaydin}
\homepage{http://www.Second.institution.edu/~Charlie.Author}
\affiliation{Institute for International Strategy, Tokyo International University, 4‐42‐31 Higashi‐Ikebukuro, Toshima‐ku, Tokyo 170‐0013, Japan}
\affiliation{Nanoelectronics Research Center, Kosuyolu Mah., Lambaci Sok., Kosuyolu Sit., No:9E/3 Kadikoy, Istanbul, T\"urkiye}

\begin{abstract}
Munero et al. developed one parameter family of mixed states $\rho^{l}$, which are more entangled than bipartite Werner state. The similar family of mixed states $\rho^{n}$ are developed by \L. Derkacz et al. with differed approach. Further the author extend $\rho^{n}$ to two parameter family of quantum states $\rho^{m}$ and  characterized these states in terms of Bell inequality violation against their mixedness. In the present article, we investigate the comparative dynamics of all mixed states $(\rho^{l},\rho^{n},\rho^{m})$ under the bipartite Ising Hamiltonian exposed by the external magnetic field and investigate the dynamics of quantum correlations against the  mixedness quantified by linear entropy.
\end{abstract}

                        
\maketitle


\section{Introduction}
Quantum information and computation is an emerging area of research~\cite{Ei1,Ei2}.
It has many applications in computation, quantum image processing, quantum sensing, and quantum communication~\cite{Qi1,Qi2,Qc1,Qc2}. To explore these applications, quantum correlations, such as entanglement and quantum discord~\cite{ent1,qd1,qd2}, are essential need in a given physical system, which is planned to use for any quantum application. During the time period falling between 2010 and 2022, significant development has been done in the context of quantum  dynamics of quantum correlations in varieties of pure and mixed quantum states involving bipartite as well as multidimensional Hilbert space; since then to date, the area of quantum dynamics is still fascinating. Additionally, during the same time period, considerable effort has been applied in quantifying and detecting quantum correlations in both pure and mixed quantum states across different Hilbert space dimensions \cite{ent2,ent3}. To reveal quantum applications, it is particularly interesting to study models in $2 \times 2$ dimensional Hilbert space, which can later be beneficial to understand various properties in many-body systems. It is immensely important to study the dynamical behavior of quantum correlations in bipartite quantum systems imposed under different physical conditions \cite{op1,op2}. So far, we have developed quite mature mathematical tools to quantify quantum correlations in bipartite systems; however, quantification challenges still persist in many-body systems. The quantum state in a bipartite system is represented by a density matrix, which needs parametrization to classify quantum states effectively. There is an ongoing demand to explore density matrices in bipartite as well as for many body systems representing pure and mixed states that have fewer parameters but exhibit a higher degree of quantum correlations~\cite{par1,bell3}.

In 2010, Tu and Ebely discovered the phenomenon of entanglement sudden death~\cite{esd1,esd2}, which states that entanglement may die in a particular quantum state for a certain time interval, then the state can not be used for any quantum application. This phenomenon has been studied in various physical settings in both bipartite and many body systems. In general, it is found that, in the absence of entanglement the quantum discord keeps its presence~\cite{esd3,esd4,esd5}, although its application has not yet been explored except quantum cryptography~\cite{qdcry}.  It is easy to calculate quantum discord in smaller systems but difficult to explore the calculations in higher dimensional Hilbert spaces; in fact, the computation of quantum discord is NP-complete~\cite{NP}. 

The study of the dynamics of quantum correlations in pure and mixed states is a necessary requirement in quantum information. It is to be noted that quantum systems are constantly interacting with external environments, making them highly elusive; pure quantum states often transform into mixed states and may lose quantum correlations. Thus, it is always interesting to investigate the structure of mixed quantum states which have maximum amount of quantum correlations~\cite{max1,max2}. In this context, Munero et al. developed a one parameter family of maximally entangled mixed states  $\rho^{l}$~\cite{mun1}, which carry more amount of entanglement than mixed Werner states in a $2\times2$ Hilbert space~\cite{w1,w2}. Subsequently, \L . Derkacz et al. developed other parameterized maximally entangled mixed states, $\rho^{n}$, which are unitary equivalent to $\rho^{l}$~\cite{ld1}. The family $\rho^{n}$ is extended to two parameter family of mixed states $\rho^{m}$, which are characterized in the context of Bell inequality violations with respect to their mixedness~\cite{bell1,mun2,bell2, bell4}. Here we mention that, the dynamical behavior of $\rho^{m}$ has not yet been explored  in literature. In the present article, we explore comparative dynamical
study of quantum correlations in the mixed quantum states $(\rho^{l},\rho^{n},\rho^{m})$ under the XX bipartite Ising Hamiltonian exposed by external magnetic field~\cite{is1,is2}. 

In section \ref{sec:mixedstateparameter}, we provide the one parameter mixed quantum states developed by Munero et al. and the similar states developed by \L . Derkacz.
We show that these states are unitary equivalent and investigate the corrosponding unitary transformation. Furthermore, we discuss their properties of Bell inequality violation with respect to their degree of mixedness. In Section \ref{sec:QCQuantification}, we provide the quantification of entanglement and quantum discord. In section \ref{sec:QCmixedstates}, a comparative dynamical study of quantum correlations is explored under XX Ising Hamiltonian with external magnetic field. In the last section, we provide the conclusion of the work.


\section{One parameter family of mixed states $(\rho^{l},\rho^{n})$}
\label{sec:mixedstateparameter}
In this section, we provide one parameter family of mixed states, $(\rho^{l},\rho^{n})$, and investigate their structure. Furthermore, we discuss the mixedness in these quantum states by using linear entropy~\cite{le1} and extend the one parameter family $\rho^{n}$ to a two parameter family $\rho^{m}$. 


\subsection{Mixed quantum states $\rho^{l}$}

Munero et al. developed a bipartite mixed quantum state ansatz in~\cite{mun1} which
is more entangled than the Werner state~\cite{w1}. This ansatz is expressed
as
\begin{equation}
\rho^{l}=\begin{pmatrix}x+\frac{\gamma}{2} & 0 & 0 & \frac{\gamma}{2}\\
0 & a & 0 & 0\\
0 & 0 & b & 0\\
\frac{\gamma}{2} & 0 & 0 & y+\frac{\gamma}{2}
\end{pmatrix}\label{eq:anszt1}
\end{equation}
with $x,y,a,b,\gamma\geq 0$ and $Tr(\rho^{l})=1$. The above ansatz can be divided into two sub categories as follows: 
\begin{equation}
\rho^{l}=
\begin{cases}
    \rho^{l}_{1}; & 
    \frac{2}{3}\leq \gamma \leq 1,
    \\
    \rho^{l}_{2}; & 
    0 \leq \gamma \leq \frac{2}{3},
  \end{cases}
\end{equation}
where,
\begin{equation}
\rho_{1}^{l}=\begin{pmatrix}\frac{\gamma}{2} & 0 & 0 & \frac{\gamma}{2}\\
0 & 1-\gamma & 0 & 0\\
0 & 0 & 0 & 0\\
\frac{\gamma}{2} & 0 & 0 & \frac{\gamma}{2}
\end{pmatrix},\quad\rho_{2}^{l}=\begin{pmatrix}\frac{1}{3} & 0 & 0 & \frac{\gamma}{2}\\
0 & \frac{1}{3} & 0 & 0\\
0 & 0 & 0 & 0\\
\frac{\gamma}{2} & 0 & 0 & \frac{\gamma}{2}
\end{pmatrix}.\label{eq:m1}
\end{equation}
On the other hand we have Werner states, which reads as follows:
\begin{eqnarray}
\rho^{W}=\gamma |\phi^{+}\rangle\langle \phi^{+}|+\frac{1-\gamma}{4}.I
\end{eqnarray}
where $I$ is the identity matrix and $|\phi^{+}\rangle$ is the singlet state. If $\gamma=0$, the state $\rho^{W}$ takes the diagonal form and hence transformed into maximally mixed state; which means it become highly incoherent. Here we mention that the ansatz by Munero et al. has linearly increasing higher amount of entanglement then the Werner state, although the maximum entanglement in both the quantum states is carried out as 1 ebit.

\subsection{Mixed quantum states $\rho^{n}$}

In this subsection, we provide a one parameter family of mixed quantum states, $\rho^{n}$, developed by L. Derkacz et al. in the context of Bell-CHSH inequality violation. For further details, we sugget the reader refer to \cite{ld1} and the references therein. Given a bipartite density matrix $\rho$, the following criteria is used to test the Bell-CHSH inequality:
\begin{equation}
m(\rho)=\bigg(2C^{2}(\rho),\lambda+C^{2\lambda}(\rho)\bigg),
\end{equation}
with $\lambda=(1-2\rho_{44})^{2}$, 
where $\rho_{44}$ is the element in the matrix $\rho$ lying at the $4^{th}$ column and $4^{th}$ row and $C(\rho)$ is the concurrence of any arbitrary quantum state
$\rho$. If $m(\rho)>1$, then the Bell-CHSH inequality
violation takes place. As per the above expression, all the states
$\rho\in\chi_{0}$, which satisfy $C(\rho)>\frac{1}{\sqrt{2}}$ violate the Bell inequality. Furthermore, the trade-off between mixedness, entanglement and Bell-CHSH inequality violation is studied by posing the problem, such that there exist a set of quantum states $\chi_{\rho}$, which have equal amount of mixedness and entanglement. Among these states, some violate Bell-CHSH inequality and some don't. The formulation is given by
\begin{equation}
\chi_{\rho}= \Bigl\{(S_{L}(\rho),C(\rho));C(\rho)>0\text{ and \ensuremath{\rho\in\chi_{\rho}}} \Bigl\},\label{eq:fam1}
\end{equation}
where mixedness is measured by the linear entropy given by the expression. 
\begin{equation}
S_{l}=\text{\ensuremath{\frac{N}{N-1}\big(1-Tr(\rho^{2})\big)}}\label{eq:le1}
\end{equation}
Following the criteria given in Eq.~$\text{\ref{eq:fam1}}$, authors developed a family of mixed quantum states in ~\cite{ld1} with the condition,
\begin{equation}
\chi_{\rho}=\{(s,c)\in R^{2}:0<c<1,0\leq s\leq S_{max}(c)\}
\end{equation}
with,
\begin{equation}
S_{max}(c)=
\begin{cases}
    \frac{8}{9}-\frac{2}{3}c^{2}; & 
    c<\frac{2}{3},
    \\
    \frac{8}{3}c(1-c); & 
    c\geq \frac{2}{3}.
  \end{cases}
\end{equation}
Authors developed the one parameter family of quantum states given as below:

\begin{widetext}
\begin{eqnarray}
\rho^{n}=\begin{pmatrix}0 & 0 & 0 & 0\\
0 & a & \frac{1}{2}ce^{\iota\theta} & 0\\
0 & \frac{1}{2}ce^{-\iota\theta} & b & 0\\
0 & 0 & 0 & 1-a-b
\end{pmatrix},\quad \mbox{with }(a,b)>0,\;\theta\in(0,2\pi). \nonumber
\end{eqnarray}
The above family $\rho^n$ can be fragmented with the range $0<c<1$ as follows: 

\begin{eqnarray}\label{eq:fm2}
\rho_{1}^{n}&=\begin{pmatrix}0 & 0 & 0 & 0\\
0 & \frac{1}{3} & \frac{1}{2}ce^{\iota\theta} & 0\\
0 & \frac{1}{2}ce^{-\iota\theta} & \frac{1}{3} & 0\\
0 & 0 & 0 & \frac{1}{3}
\end{pmatrix},\; \mbox{and } \rho_{2}^{n}=\begin{pmatrix}0 & 0 & 0 & 0\\
0 & \frac{c}{2} & \frac{1}{2}ce^{\iota\theta} & 0\\
0 & \frac{1}{2}ce^{-\iota\theta} & \frac{c}{2} & 0\\
0 & 0 & 0 & 1-c
\end{pmatrix},
\end{eqnarray}
\end{widetext}

where $\rho_1^n$ is defined for the range $c\in(0,\frac{2}{3})$, while $\rho_2^n$ is defined for $c\in[\frac{2}{3},1)$. It is to be noted that the above quantum states $(\rho_{1}^{n},\rho_{2}^{n})$
with $(\theta=0, c=\gamma)$ are equivalent to ($\rho_{1}^{l},\rho_{2}^{l})$
respectively under a unitary transformation and hence they carry
same amount of quantum correlations. Here we develop the unitary transformation satisfying $U\rho_{1}^{l}U^{\dagger}=\rho_{2}^{n}$ and $U\rho_{2}^{l}U^{\dagger}=\rho_{1}^{n}$, which is given by
\begin{eqnarray}
U=\begin{pmatrix}0 & 0 & 1 & 0\\
0 & 0 & 0 & 1\\
1 & 0 & 0 & 0\\
0 & 1 & 0 & 0
\end{pmatrix}
\end{eqnarray}

\subsection{Two parameter family of mixed states $\rho^{m}$}

Following the calculations provided in~\cite{ld1},
author extended the one parameter family of quantum states, $\rho^{n}$,
to the two parameter family mixed quantum states, $\rho^{m}$, under the conditions $C(\rho)\leq\frac{1}{\sqrt{2}}$ and $0\leq s\leq1$ belonging to concurrence and nixedness respecively. Furthermore, a new parameter $D$ is introduced,  which is expressed in terms of concurrence and linear entropy as follows:

\begin{widetext}
\begin{equation}
D=-\frac{c^{2}}{12}-\frac{s}{8}+\frac{1}{9}\label{eq:d1}
\end{equation}
The two parameter mixed states are obtained as given below: 
\begin{eqnarray}
\rho^{m}=\begin{pmatrix}0 & 0 & 0 & 0\\
0 & \frac{1}{3}+\sqrt{D}(\sin\phi+\sqrt{3}\cos\phi) & \frac{c}{2}e^{\iota\theta} & 0\\
0 & \frac{c}{2}e^{-\iota\theta} & \frac{1}{3}+\sqrt{D}(\sin\phi-\sqrt{3}\cos\phi) & 0\\
0 & 0 & 0 & \frac{1}{3}-2\sqrt{D}\sin\phi
\end{pmatrix}\label{eq:fm3}
\end{eqnarray}
\end{widetext}
with $\phi\in[0,2\pi].$ The authors categorized the above family of quantum
states with three different conditions, with the fixed value of concurrence $(c=\frac{1}{2})$ and varying linear entropy as $(S=\frac{1}{8},S=\frac{1}{2},S=\frac{7}{10})$. For the combinations $(s,c)=\{(\frac{1}{8},\frac{1}{2}),(\frac{1}{2},\frac{1}{2}),(\frac{7}{10},\frac{1}{2})\}$ the corresponding values of the parameter $D$ are obtained as $(D=\frac{\sqrt{43}}{24},D=\frac{1}{6},D=\frac{1}{6\sqrt{10}})$. By putting the values of the parameter $D$ and $c$ in Eq.~\ref{eq:fm3}, the three versions of the quantum states are obtained, which are discussed in subsequent subsections.

\subsubsection{Mixed quantum states $\rho_{1}^{m}$}
Substituting the parameter values $\sqrt{D}=\frac{\sqrt{43}}{24},\theta=0,c=\frac{1}{2}$ in Eq.~\ref{eq:fm3}, the following family of quantum states is obtained, which violate the Bell-CHSH inequality, satisfying $m(\rho)>1$ with the parameter range
$\phi\in(0.54657,0.65605)$. 
\begin{widetext}
\begin{equation}
\rho_{1}^{m}=\begin{pmatrix}0 & 0 & 0 & 0\\
0 & \frac{1}{3}+\frac{\sqrt{43}}{24}(\sin\phi+\sqrt{3}\cos\phi) & \frac{1}{4} & 0\\
0 & \frac{1}{4} & \frac{1}{3}+\frac{\sqrt{43}}{24}(\sin\phi-\sqrt{3}\cos\phi) & 0\\
0 & 0 & 0 & \frac{1}{3}-\frac{\sqrt{43}}{12}\sin\phi
\end{pmatrix}\label{eq:fm4}\end{equation}
\end{widetext}
\subsubsection{Mixed quantum states $\rho_{2}^{m}$}
Substituting the parameter values $\sqrt{D}=\frac{1}{6},\theta=0,c=\frac{1}{2}$ in Eq.~\ref{eq:fm3}, a new family of quantum states is obtained, satisfying
$-1<m(\rho)<1$ within the range $\phi\in(0.25,1.57)$. These quantum states do not violate the Bell-CHSH inequality for $\phi\in(0.25,0.9)$, on the other hand the  Bell-CHSH inequality violation take place with $\phi\in(0.9,1.57)$. The states reads as:
\begin{widetext}
\begin{equation}
\rho_{2}^{m}=\begin{pmatrix}0 & 0 & 0 & 0\\
0 & \frac{1}{3}+\frac{1}{6}(\sin\phi+\sqrt{3}\cos\phi) & \frac{1}{4} & 0\\
0 & \frac{1}{4} & \frac{1}{3}+\frac{1}{6}(\sin\phi-\sqrt{3}\cos\phi) & 0\\
0 & 0 & 0 & \frac{1}{3}-\frac{1}{3}\sin\phi
\end{pmatrix}\label{eq:fm5}
\end{equation}
\end{widetext}
\subsubsection{Mixed quantum states $\rho_{3}^{m}$}
Substituting the parameter values $\sqrt{D}=\frac{1}{6\sqrt{10}},\theta=0,c=\frac{1}{2}$ in Eq.~\ref{eq:fm3}, the following family of quantum states is obtained, satisfying $m(\rho)<1$. These quantum states do not violate Bell-CHSH inequality for the parameter $\phi\in(0,2\pi)$. These states read as:
\begin{widetext}
\begin{eqnarray}
\rho_{3}^{m}=\begin{pmatrix}0 & 0 & 0 & 0\\
0 & \frac{1}{3}+\frac{1}{6\sqrt{10}}(\sin\phi+\sqrt{3}\cos\phi) & \frac{1}{4} & 0\\
0 & \frac{1}{4} & \frac{1}{3}+\frac{1}{6\sqrt{10}}(\sin\phi-\sqrt{3}\cos\phi) & 0\\
0 & 0 & 0 & \frac{1}{3}-\frac{1}{3\sqrt{10}}\sin\phi
\end{pmatrix}\label{eq:fm6}
\end{eqnarray}
\end{widetext}

\section{Quantification of Quantum correlations}
\label{sec:QCQuantification}
In this section, we outline the methods used for quantifying quantum correlations in the present work. For entanglement quantification, we have used Wooters bipartite concurrence~\cite{con1,con2}, while quantum discord~\cite{qd1,qd2} is used to quantify the measurement based quantum correlations. Furthermore, to quantify the mixedness of quantum states we have used linear entropy given in Eq.\ref{eq:le1}.

\subsection{Concurrence}

The entanglement in these states can be calculated by using the concurrence.
The concurrence $C(\rho)$ in a given density matrix $\rho$ is given
by 
\begin{equation}
C(\rho)=\text{max\{0,\ensuremath{\lambda_{1}-\lambda_{2}-\lambda_{3}-\lambda_{4}}\}},
\end{equation}
where $\lambda_{i}$ with ($i=1,2,3,4$) are the square root of the
eigenvalues in decreasing order of $\rho\rho^{f}$, where $\rho^{f}$is
the spin flip density matrix given as,

\begin{equation}
\rho^{f}=(\sigma^{y}\otimes\sigma^{y})\rho^{\star}(\sigma^{y}\otimes\sigma^{y})\label{eq:_con}
\end{equation}
Where $\rho^{\star}$ is the complex conjugate of the density matrix
and $\sigma^{y}$ are the Pauli Y operator. The combination of the
matrix $\rho\rho^{f}$ is used to calculate the concurrence in a given
density matrix, it is to be noted that concurrence is unitary invariant.
\begin{figure*}
\centering
\includegraphics[width=0.9\textwidth]{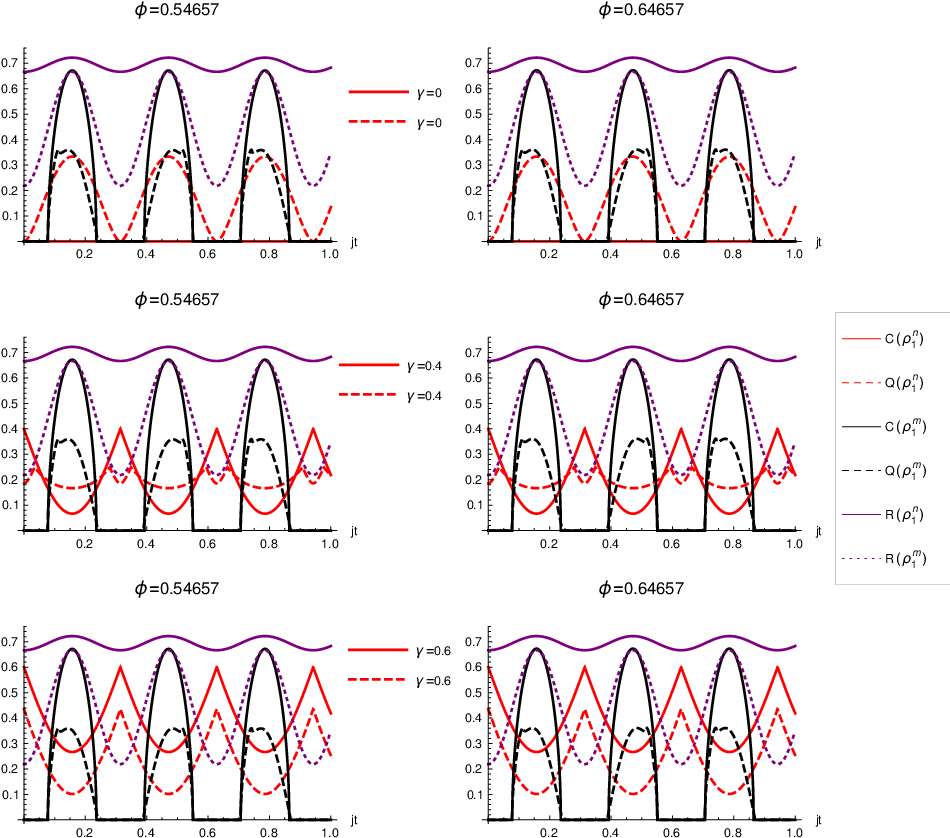}
\caption{Entanglement, Quantum Discord and Linear Entropy in $(\rho_{1}^{n},\rho_{1}^{m})$}
\label{fig:1}      
\end{figure*}

\subsection{Quantum discord}
Quantum discord is a quantum measurement based quantum correlation
that can be quantified as the difference between mutual information
and classical correlation in bipartite systems. In general mutual
quantum information is not equal to the classical mutual information,
since the quantum mutual information quantification depends on the
measurement carried on sub party involved in bipartite system. It
is very important to note that quantum discord is not symmetric with
respect to quantum measurement. The expression of quantum discord
is given as below,

\begin{equation}
Q(\rho^{AB})=I(\rho^{AB})-C(\rho^{AB})
\end{equation}
which can be re-expressed as,

\begin{equation}
Q(\rho^{AB})=S(\rho^{A})-S(\rho^{AB})+\underset{\{A_{k}\}}{\text{min}}S(\rho^{B}|A_{k})\label{eq:_qd}
\end{equation}
The last term denotes the minimization taken over the entropy in reduced
density matrix $\rho^{B}$ with the set of measurement operators $\{A_{k}\}$. The above expression is
optimization problem and difficult to solve except the density matrix is in the from of X states. However calculating quantum discord is NP complete\cite{NP}.


\section{Quantum correlations dynamics in mixed states}
\label{sec:QCmixedstates}
This section is divided in two subsections. In subsection \ref{4.1}, we study the quantum correlations and mixedness in all the mixed quantum states $(\rho_{1}^{m},\rho_{2}^{m},\rho_{3}^{m})$ with their respective  range of the parameter $\phi$. Here we mention that all these states are defined with $C(\rho)<\frac{1}{\sqrt{2}}$, which leads $0<C(\rho)<0.70710678118$, further the states $\rho_{1}^{n}$ are defined with $c\in(0,\frac{2}{3})$, which leads $0<C(\rho_{1}^{n})<0.6\Bar{6}$. So it is quite logical to study the comparative dynamics of $(\rho_{1}^{m},\rho_{2}^{m},\rho_{3}^{m})$ with $\rho_{1}^{n}$. In subsection \ref{4.3} and onwards, the comparative dynamics of quantum correlations under the XX Ising Hamiltonian exposed by external magnetic field has been explored.

\subsection{Quantum correlations in mixed states}\label{4.1}
Here we provide the amount of entanglement, quantum discord and linear entropy carried out for the mixed states  $(\rho_{1}^{m},\rho_{2}^{m},\rho_{3}^{m})$ respectively. 

\subsection{Quantum correlations under Ising Hamiltonian}
Under this section we study the time evolution of quantum correlations
by using the Ising Hamiltonian exposed by the external magnetic field.
Here we consider the XX interaction, the Hamiltonian of the systems
is given below,
\begin{eqnarray}
H(\sigma) & = & \sum_{<i,j>}J_{ij}\sigma_{i}\sigma_{j}+B_{z}.\sigma_{z}
\end{eqnarray}
We have studied the dynamics in Schrodinger picture by considering
the unitary time evolution operator $U(t)=U^{-iHt/\hbar}$.
We obtain the time evolution of the density matrix as, 
\begin{equation}
\rho(t)=U(t)\rho U^{\dagger}(t)\label{eq:te1}
\end{equation}

\begin{figure*}
\centering
\includegraphics[width=0.9\textwidth]{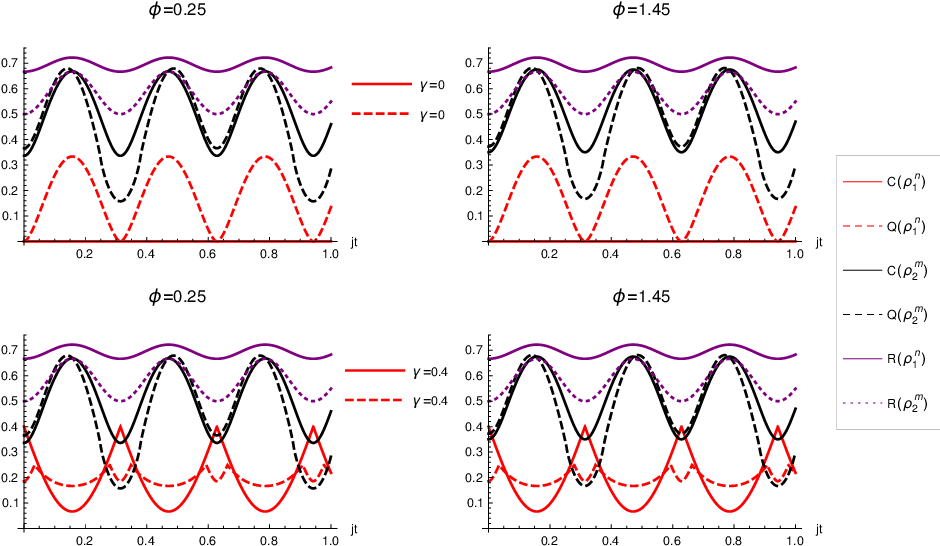}
\caption{Entanglement, Quantum Discord and Linear Entropy in $(\rho_{1}^{n},\rho_{2}^{m})$}
\label{fig:2}      
\end{figure*}

\subsection{Comparative dynamics in $(\rho_{1}^{n}, \rho_{1}^{m})$}\label{4.3}
We consider the quantum states $\rho_{1}^{n}$ given in 
Eq.~$\text{\ref{eq:fm2}}$ and recall that the states $\rho_{1}^{n}$ are unitary equivalent to 
$\rho_{2}^{l}$. The time evolution in the density matrix $\rho_{1}^{n}$ and the linear entropy are obtained as below,
\begin{equation}
\rho_{1}^{n}(t)=\left(\begin{array}{cccc}
\frac{1}{3} & 0 & 0 & \frac{1}{2}\gamma e^{-4\iota Bt}\\
0 & \frac{\cos^{2}(\text{Jt})}{3} & \frac{1}{6}\iota\sin(2\text{Jt}) & 0\\
0 & -\frac{1}{6}\iota\sin(2\text{Jt}) & \frac{\sin^{2}(\text{Jt})}{3} & 0\\
\frac{1}{2}\gamma e^{4\iota Bt} & 0 & 0 & \frac{1}{3}
\end{array}\right)
\label{eq:_time_Evo_m}
\end{equation}

\begin{equation}
S_{1}^{n}(t)=\frac{1}{36}(25-\cos(4\text{Jt}))
\label{eq:_le_m}
\end{equation}
Here we mention a fact that the density matrix $\rho_{1}^{n}(t)$ involve the term of magnetic field $B$, this term is vanished while calculating the quantum correlations in the density matrix in further sections. Next, the time evolution density matrix and linear entropy corresponding
to the state $\rho_{1}^{m}$ is given below,

\begin{eqnarray}
\rho_{1}^{m}(t)&=&\left(\begin{array}{cccc}
0 & 0 & 0 & 0\\
0 & \frac{1}{24}\left(x+y+8\right) & \frac{1}{24}\left(6+\iota x\right) & 0\\
0 & \frac{1}{24}\left(6-\iota x\right) & \frac{1}{24}\left(-x+y+8\right) & 0\\
0 & 0 & 0 & \frac{1}{12}\left(4-y\right)
\end{array}\right)\label{eq:_rho_t_n1} \\
\mbox{with, } &\\ \nonumber
x&=&\sqrt{129}\cos(2\text{Jt})\cos(\phi), \quad
y=\sqrt{43}\sin(\phi).
\end{eqnarray}

Next we have obtained the linear entropy in the same state, which
reads as,
\begin{equation}
S_{1}^{m}(t)=\frac{1}{96}\left(-43\cos^{2}(2\text{Jt})\cos^{2}(\phi)-43\sin^{2}(\phi)+64\right)\label{eq:_Le_n}
\end{equation}
The Concurrence and quantum discord are simulated by using the equations
Eq.~$\text{\ref{eq:_con}}$ and Eq.~$\text{\ref{eq:_qd}}$ respectively. The simulated results of concurrence, quantum discoed and mixedness vs. the parameter
$Jt$ is shown in the Fig.1. The solid red and
dotted red color represents concurrence and quantum discoed in $\rho_{1}^{n}$ respectively; while the solid black and
dotted black colors represents concurrence and quantum discoed in $\rho_{1}^{m}$ respectively. The linear entropy is represented in both the states with solid violate and dotted violate colors respectively. The same color code and style is followed throughout the paper for every figure.
We have found the quantum corrections and mixedness are not affected
by the external magnetic filed in both the states $\rho_{1}^{n}$ and $\rho_{1}^{m}$ for any value of $(\gamma,\phi)$. For $\gamma=0$, it is observed that the entanglement is zero $\rho_{1}^{n}$, while periodic quantum discord exists. On the other hand the periodic quantum correlations sudden death has been observed in the state $\rho_{1}^{m}$ with  $(\phi=0.54657,\phi=0.65605)$ as the parameter $Jt$ advances. However the periodic amplitude of the entanglement in $\rho_{1}^{m}$ is less then the amplitude of quantum discord.    
As the value of $\gamma=0.4$ is achieved with $(\phi=0.54657,0.64657)$, the initial amplitude of entanglement and quantum discord are lifted in $\rho_{1}^{n}$ from zero to $0.2$ and $0.4$ respectively. As the parameter $Jt$ advances, both quantum correlations exhibit periodic behavior  and entanglement is always higher then the quantum discord in both the quantum states. The advancement in the amount of $\gamma$ i.e. $\gamma=0.6$ with $\phi=(0.54657,0.64657)$ leads more amplification of quantum correlations in $\rho_{1}^{n}$, but still the behavior of quantum correlations freezes in the quantum state $\rho_{1}^{m}$. It is noted that the peak value of periodic quantum discord in $\rho_{1}^{m}$ is less then $\rho_{1}^{n}$.
In both the quantum states the behavior of linear entropy is periodic and degree of mixedness in $\rho_{1}^{n}$ is higher then $\rho_{1}^{m}$ as $Jt$ advances.  

\begin{figure*}
\centering
\includegraphics[width=0.9\textwidth]{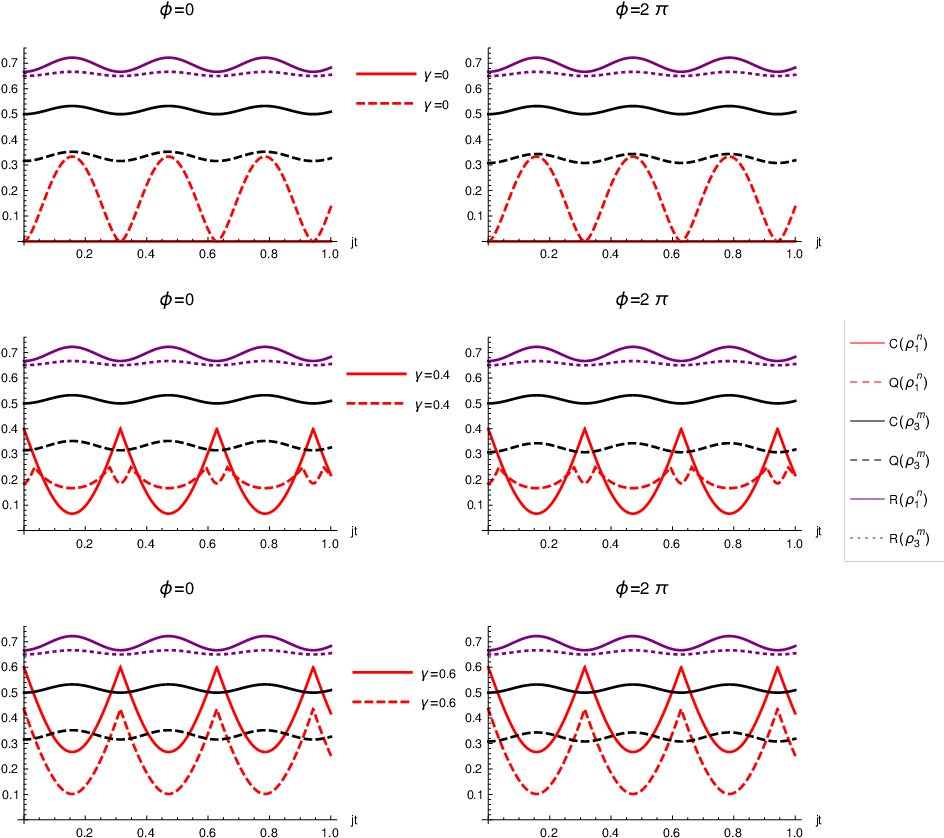}
\caption{Entanglement, Quantum Discord and Linear Entropy in $(\rho_{1}^{n},\rho_{3}^{m})$}
\label{fig:3}      
\end{figure*}

\subsection{Comparative dynamics in $(\rho_{1}^{n},\rho_{2}^{m})$}

In this subsection we provide the comparative dynamics of quantum
correlations and mixedness in the quantum states $\rho_{1}^{n}$ and
$\rho_{2}^{m}$ respectively. The time evolution density matrix and linear entropy in $\rho_{1}^{n}$ is obtained in 
Eq.~\ref{eq:_time_Evo_m} and
\ref{eq:_le_m}; while the time evolution density matrix and linear entropy in $\rho_{2}^{m}$ is given below,
\begin{equation}
\rho_{2}^{m}(t)=\left(
\begin{array}{cccc}
 0 & 0 & 0 & 0 \\
 0 & \frac{1}{3}+a & \frac{1}{4}+b & 0 \\
 0 & \frac{1}{4}-b & \frac{1}{3}-a & 0 \\
 0 & 0 & 0 & 0 \\
\end{array}
\right)
\end{equation}
with,
\begin{eqnarray}
a=\frac{\sqrt{3}}{6}\cos(2Jt)\cos(\phi)+\sin(\phi), \quad
b=\frac{\sqrt{3}\iota}{6}\cos(\phi)+\sin(2Jt)\nonumber
\end{eqnarray}
Linear entropy is given by,
\begin{eqnarray}
S_{2}^{m}(t)=\frac{1}{6} \left(-\cos ^2(2 \text{Jt}) \cos ^2(\phi)-\sin ^2(\phi)+4\right)
\end{eqnarray}
The graphical results with the parameter range 
$\phi\in(0.25,1.45)$ is plotted in Fig.2. For $\gamma=0$ with $\phi=(0.25,1.45)$, there is no entanglement in $\rho_{1}^{n}$, but periodic quantum correlations takes place in state. As the value of the parameter $\gamma$ increases, the quantum correlations in the state $\rho_{1}^{n}$ exhibit the same behavior as discussed in sub-section \ref{4.3}. Although with the advancement of the parameter $\phi$, the quantum state $\rho_{2}^{m}$ exhibit more robust behavior then $\rho_{1}^{n}$; whenever the quantum correlations are absent in $\rho_{1}^{n}$, the quantum correlations exists in $\rho_{2}^{m}$. It is observed that the quantum discord in $\rho_{2}^{m}$ is higher then entanglement in $\rho_{2}^{m}$ except peak values; however the amount of peak values of quantum correlations are same as $0.65$.

For $\gamma=0.4$, the initial amplitude of entanglement and quantum discord in $\rho_{1}^{n}$ at $Jt=0$ is lifted from zero to $0.4$. The peak values of quantum correlations in $\rho_{2}^{m}$ is higher then $\rho_{1}^{n}$, hence in this context also the state is more robust then $\rho_{1}^{n}$. In both the state the linear entropy exhibits the periodic behavior, the peak values of the linear entropy is synchronized with the peak values of the quantum correlations in the state $\rho_{2}^{m}$, but the degree of mixedness in $\rho_{1}^{n}$ is higher then $\rho_{2}^{m}$. More specifically we have found that the quantum state $\rho_{2}^{m}$ does not suffer from sudden death of quantum correlations and also exhibit a robust character then $\rho_{1}^{m}$. 

\subsection{Comparative dynamics in $(\rho_{1}^{n},\rho_{3}^{m})$}
In this subsection we explore the results of the dynamics of quantum
correlations in $\rho_{1}^{n}$ and $\rho_{3}^{m}$ for different
values of $\gamma$ with the range $(\phi=2\pi)$. We recall that, the time evolution density matrix and linear entropy in $\rho_{1}^{n}$ is obtained in 
Eq.~\ref{eq:_time_Evo_m} and
\ref{eq:_le_m} respectively; further we provide the time evolution density matrix and linear entropy in $\rho_{3}^{m}$ as below,
\begin{equation}
\rho_{3}^{m}(t)=\left(
\begin{array}{cccc}
 0 & 0 & 0 & 0 \\
 0 & \frac{1}{3}+g & \frac{1}{4}+h & 0 \\
 0 & \frac{1}{4}-h & \frac{1}{3}-g & 0 \\
 0 & 0 & 0 & 0 \\
\end{array}
\right)
\end{equation}
with,
\begin{eqnarray}
g=\frac{\sqrt{30}}{60}\cos(2Jt)\cos(\phi)+\frac{\sqrt{10}}{60}\sin{\phi}, \quad
h=\frac{\iota \cos(\phi)\sin(2Jt)}{2\sqrt{30}} \nonumber 
\end{eqnarray}
The linear entropy reads as,
\begin{equation}
 S_{3}^{m}(t)=\frac{1}{60} \left(-\cos ^2(2 \text{Jt}) \cos ^2(\phi  )-\sin ^2(\phi+40\right)   
\end{equation}
We plotted the graphical results of quantum correlations and linear entropy in both the quantum states $(\rho_{1}^{n},\rho_{3}^{m})$ in Fig 3. with the parameter range $\phi\in(0,2\pi)$. For $\gamma=0$, the behavior of quantum correlations in $\rho_{1}^{n}$ remains same as it is discussed in sub-section \ref{4.3}. On the other hand the peak values of periodic quantum correlations in $\rho_{3}^{m}$ is always higher then $\rho_{1}^{n}$.

For $\gamma=0.4$, the peak value of entanglement in $\rho_{1}^{n}$ overshoots the amplitude of quantum discord in $\rho_{3}^{m}$. Further as the value of $\gamma$ reaches to $0.6$, the periodic peak values of both entanglement and quantum discord overshoot the peak amplitudes of both the quantum correlations in $\rho_{3}^{m}$;  so in the context of peak values the quantum state $\rho_{1}^{n}$ exhibit robust character in comparison to the quantum state $\rho_{3}^{m}$. On the other hand the amount of entanglement in $\rho_{3}^{m}$ is higher then quantum discord. However the linear entropy in both the states also shows periodic behavior and degree of mixedness in $\rho_{1}^{n}$ is higher then $\rho_{3}^{m}$. 


\section{Conclusion}
In this article we have studied the comparative dynamics of quantum correlations
and mixedness in one parameter quantum states developed by Munero et al. i.e. $\rho_{1}^{n}$
and developed by \L . Derkacz et al. i.e $(\rho_{1}^{m},\rho_{2}^{m},\rho_{3}^{m})$ under the XX Ising Hamiltonian
exposed by external magnetic field. All these states falls in the category of X states, we have found that all the states never affected by external magnetic field while quantum correlations dynamics is obtained. 

We have found interesting results that the states $\rho_{1}^{m}$ suffer from the sudden
death of quantum correlations, hence in this context the state does not exhibit the robust character against the state $\rho_{1}^{n}$. While on the other hand the quantum states $(\rho_{2}^{m},\rho_{3}^{m})$ never suffer from quantum correlations sudden death and have robust character then the quantum state $\rho_{1}^{m}$. As the peak values of quantum correlations are concern, the increasing value of $\gamma$ in $\rho_{1}^{n}$ overshoots the peak amplitudes of quantum correlations in $(\rho_{2}^{m},\rho_{3}^{m})$; and hence in this context the quantum state $\rho_{1}^{n}$ exhibit more robust character then these states. Further we have found that the degree of periodic mixedness in $\rho_{1}^{n}$ is always higher then all the states $(\rho_{1}^{m},\rho_{2}^{m},\rho_{3}^{m})$. The present work incorporating comparative dynamics of the quantum correlations in all the quantum states $(\rho_{1}^{m},\rho_{2}^{m},\rho_{3}^{m})$ with $\rho_{1}^{n}$ can be beneficial for quantum information community.

%


\end{document}